\documentstyle [12pt] {article}

\title{Vacuum state of the quantum string without anomalies in
any number of dimensions
\thanks{\it{This work was partially supported by  Comision de
Investigaciones Cientificas, Pcia. de Buenos Aires; Argentina.}}}
\author{C.G.Bollini, M.C.Rocca\\
Departmento de Fisica, Fac. de Ciencias Exactas,\\
Universidad Nacional de La Plata.\\
C.C. 67 (1900) La Plata. Argentina}
\date{June 1, 1996}

\begin{document}

\maketitle
\begin{abstract}
 We show that the anomalies of the Virasoro algebra are due to the
asymmetric behavior of raising and lowering operators with respect to
the ground state of the string. With the adoption of a symmetric
vacuum we obtain a non-anomalous theory in any number of dimensions.
In particular for D=4.


\end{abstract}

\eject

\section {Introduction}
 String theory has been a source of many excellent works on the subject.
Some of them give rise to the hope that a "unified and final"
description of all forces of nature can be achieved.

 However the appearance of anomalies seems to impose a severe restriction
on the number of dimensions of the spaces in which such a program can be
succesfully carried out.

 In reference \cite{tp1} a didactic exposition of string theory can be
found. We will adopt this book as our reference text on the subject.
See also references \cite{tp2} and \cite{tp3}.

 The position of the string in space time is given by the coordinates
$ {X}^{\mu} (\sigma , \tau) $. The action can be written:

\begin{equation}
S=- \frac {1} {2 \pi} \int d^2 \sigma \;\; {\eta}^{\alpha \beta}\;\;
{\partial}_{\alpha} X \cdot {\partial}_{\beta} X
\end{equation}
where $ {\eta}^{\alpha \beta} $ is a two dimensional Minkowsky metric
tensor.

 The action (1) leads to the wave equation

\begin{equation}
\left( \frac {{\partial}^2} {{\partial \sigma}^2} -
\frac {{\partial}^2} {{\partial \tau}^2} \right) X^{\mu} =0
\end{equation}

 Its general solution is a superposition of a right moving coordinate
$ X_R^{\mu} $ and a left moving one $ X_L^{\mu} $ . For the closed
string \cite{tp4} :

\begin{equation}
X_R^{\mu} = \frac {1} {2} x^{\mu} + \frac {1} {2} p^{\mu} (\tau
- \sigma) + \frac {i} {2} \sum\limits_{n \neq 0 } \frac {1} {n}
{\alpha}_n^{\mu} e^{-2in \left( \tau - \sigma \right)}
\end{equation}

\begin{equation}
X_L^{\mu} = \frac {1} {2} x^{\mu} + \frac {1} {2} p^{\mu} (\tau
+ \sigma) + \frac {i} {2} \sum\limits_{n \neq 0 } \frac {1} {n}
{\tilde{\alpha}}_n^{\mu} e^{-2in \left( \tau + \sigma \right)}
\end{equation}

 For the open string we have standing waves:

\begin{equation}
X^{\mu} = x^{\mu} + p^{\mu} \tau + i \sum\limits_{n \neq 0} \frac {1} {n}
{\alpha}_n^{\mu} e^{-in \tau} \cos {n \sigma}
\end{equation}

 The Virasoro generators can be defined as the Fourier components
of $ {\dot{X}}^2 $ \cite{tp5}:

\begin{equation}
L_m = \frac {1} {2 \pi} {\int}_0^{\pi} d \sigma \;\; e^{-2im \sigma}
{\dot{X}}_R^2 = \frac {1} {2} \sum\limits_{- \infty}^{\infty}
{\alpha}_{m-n} \cdot {\alpha}_{n}
\end{equation}

\begin{equation}
{\tilde{L}}_m = \frac {1} {2 \pi} {\int}_0^{2 \pi} d \sigma
e^{-2im \sigma} {\dot{X}}_L^2 = \frac {1} {2}
\sum\limits_{- \infty}^{\infty} {\tilde{\alpha}}_{m-n} \cdot
{\tilde{\alpha}}_n
\end{equation}

(with $ {\alpha}_0^{\mu} $ $ = p^{\mu} $ ).

\section {Quantization}

 With the string action given by (1), the momentum canonically
conjugate to the coordinate $ X^{\mu} $ is proportional to
$ {\dot{X}}^{\mu} $ . Using (3) and  (4) it is possible to find
the commutation relations obeyed by the coefficients
$ {\alpha}_m^{\mu} $ \cite{tp6} :

\begin{equation}
\left[ {\alpha}_m^{\mu}, {\alpha}_n^{\nu} \right] = m
{\delta}_{m+n} {\eta}^{ \mu \nu}
\end{equation}

 Except for normalization, eq. (8) is the usual rule for the harmonic
oscilator raising and lowering operators.

 The commutation relations obeyed by the Virasoro operators can be
found by using (6) and the rules given by (8). The result is the
Virasoro algebra:

\begin{equation}
\left[ L_m, L_n \right] = (m-n) L_{m+n} \;\;,\;\; (m+n \neq 0)
\end{equation}

 When n=-m, one must be careful. A c-number central term may appear
and there are some ordering ambiguities.

 We define:

\begin{equation}
\left[ L_m, L_n \right] = (m-n) L_{m+n} + A_m {\delta}_{m+n}
\end{equation}

 Where $ A_m $ is called the anomaly of the Virasoro algebra.
If we take n=-m in (10) we have:

\begin{equation}
\left[ L_m, L_{-m} \right] = 2mL_0 + A_m
\end{equation}

 In view of eq.(6), the natural form for $ L_0 $ is:

\begin{equation}
L_0= \frac {1} {2} {\alpha}_0^2 + \frac {1} {2} \sum\limits_{n=1}^{\infty}
({\alpha}_{-n} \cdot {\alpha}_n + {\alpha}_n \cdot {\alpha}_{-n})
\end{equation}

 However, a normal ordered expression for $ L_0 $ is generally
understood. Anyway, a c-number addition to $ L_0 $ is equivalent to
a redefinition of the anomaly $ A_m $.

 If $ L_0 $ is defined so as to have a null vacuum expectation
value, then

\begin{equation}
A_m = \langle 0 \mid \left[ L_m, L_{-m} \right] \mid 0 \rangle
\end{equation}

\section {The vacuum state}

 To build-up the states of the string, it is necessary to complement
the basic commutation rules (eq.(8)), with a definition of the
vacuum.

 The usual definition takes the vacuum as the state annihilated by
all $ {\alpha}_m^{\mu} $ with $ m > 0 $.
As we are going to change this
identification for the string, some words are needed to support such
an attitude.

 There are good reasons to adopt the costumary definition. In particle
physics it leads to a positive definite spectrum for the hamiltonian.
It also leads to the Feynman propagator, which satisfies the
asymptotic requirement that the positive energies propagate towards
the future and the negative energies propagate backwards in time.

 However, for waves in a string (either open or closed), there are no
asymptotic requirements in space. These waves are confined to the
string and can not be observed as free waves. An analogous example
is provided by waves that obey a Klein-Gordon equation with
complex-mass parameter \cite{tp7}. It is not possible to see those
waves in free states, as they blow-up asymptotically. Something
similar happens with waves associated with the possible existence
of tachyons \cite{tp8} \cite{tp9}. See also ref \cite{tp10}.

 In all those cases the vacuum plays a symmetrical role with respect
to the raising and lowering operators.

 It is also possible that the ground state has been incorrectly
identified, as mentioned in ref. \cite{tp11}. In which case the
symmetrical vacuum seems  to be a reasonable alternative to the usual
definition.

 These considerations lead us to the assumption that the vacuum
state of the string should be annihilated not by $ {\alpha}_m^{\mu} $
(or $ {\alpha}_{-m}^{\mu} $ ), but rather by the symmetrized product
$ \{ {\alpha}_m^{\mu}, {\alpha}_{-m}^{\mu} \} $ (no summation over
$ \mu $ or m ).

 In other words, the vacuum obeys (See refs. \cite{tp7} to
\cite{tp10})

\begin{equation}
\left( {\alpha}_m^{\mu} {\alpha}_{-m}^{\mu} + {\alpha}_{-m}^{\mu}
{\alpha}_m^{\mu} \right) \mid 0 \rangle = 0 \;\;\;(no\;summation)
\end{equation}

 As a consequence we have

\begin{equation}
L_0 \mid 0 \rangle =0
\end{equation}

where $ L_0 $ is defined by eq.(12).

 It also follows that, for any $ \mu $, $ \nu $, m, n:

\begin{equation}
\langle 0 \mid {\alpha}_m^{\mu} {\alpha}_n^{\nu} +
{\alpha}_n^{\nu} {\alpha}_m^{\mu} \mid 0 \rangle = 0
\end{equation}

 As

\[ {\alpha}_m^{\mu} {\alpha}_n^{\nu} = \frac {1} {2} \left[
{\alpha}_m^{\mu},{\alpha}_n^{\nu} \right] + \frac {1} {2}
\lbrace {\alpha}_m^{\mu}, {\alpha}_n^{\nu} \rbrace \]

it is easy to see that :

\begin{equation}
\langle 0 \mid {\alpha}_m^{\mu} {\alpha}_n^{\nu} \mid 0 \rangle =
\frac {1} {2} m {\delta}_{m+n} {\eta}^{\mu \nu}
\end{equation}

while in the usual case:

\begin{eqnarray}
\langle 0 \mid {\alpha}_m^{\mu} {\alpha}_n^{\nu} \mid 0 \rangle & = &
m {\delta}_{m+n} {\eta}^{\mu \nu} \;\;\; if \;\; m>0 \nonumber \\
& = & 0 \;\;\; if \;\; m<0
\end{eqnarray}

We would like to point out that when the vacuum state is not
anihilated by the decreasing operator, a set of negative normed
states appears.

 For example, if we define

\[ \mid -{1}_n \rangle \equiv {\alpha}_n \mid 0 \rangle \;\;\;\;
n \rangle 0 \]

 Then, according to eq.(17):

\[ \langle -{1}_n \mid -{1}_n \rangle = \langle 0 \mid {\alpha}^{+}_n
{\alpha}_n \mid 0 \rangle = - \frac {1} {2} n \]

 And the states $ \mid -{1}_n \rangle $ (n>0) are negatively
normed.

 However, as explained in references [7] to [10] , we have to take
into account that different choices of the vacuum imply different
propagators. For the usual case one obtains Feynman's Green
function while for the symmetrical vacuum we get Wheeler's
propagator (half advanced and half retarded \cite{tp12} ).
As is well-known, Feynman's causal function has an on-shell pole,
which signals the existence of a corresponding free state. On the
other hand, Wheeler's propagator as an on-shell {\bf zero } .
Thus implying that the corresponding free mode can not be
excited. So that here the negative normed states are harmless.

 It is to be noted that, in spite of its advanced component, the
Wheeler's Green function does not give rise to causal
inconsistencies \cite{tp13} .

\section{The anomalies}

 Before engaging in an actual calculation, it seems convenient to give
an intuitive approach to the subject.
 Let us first consider the usual case. $ L_0 $ is supposed to be normal
ordered, so that eq.(13) is valid. For $ m > 0 $ we have:

\begin{equation}
{\alpha}_m^{\mu} \mid 0 \rangle = 0 \;\;\; and \;\;\; L_m \mid 0
\rangle = 0
\end{equation}

 On the other hand,

\begin{equation}
L_{-m} \mid 0 \rangle = \frac {1} {2} \sum\limits_{n=1}^{m-1}
{\alpha}_{n-m} \cdot {\alpha}_{-n} \mid 0 \rangle
\end{equation}

 So that:

\[ A_m = \langle 0 \mid \left[L_m,L_{-m} \right] \mid 0 \rangle =
\langle 0 \mid L_m L_{-m} \mid 0 \rangle \]

\[ A_m = \frac {1} {4} \sum\limits_{n=1}^{m-1} \sum\limits_{s=1}^{m-1}
\langle 0 \mid {\alpha}_{m-n} \cdot {\alpha}_n \; {\alpha}_{s-m}
\cdot {\alpha}_{-s} \mid 0 \rangle \]

\[ \;\;\; = \frac {1} {4} {\eta}_{\mu \nu} {\eta}_{\rho \sigma}
\sum\limits_{n=1}^{m-1} \sum\limits_{s=1}^{m-1} \langle 0 \mid \left(
{\alpha}_{m-n}^{\mu} \left[ {\alpha}_n^{\nu}, {\alpha}_{s-m}^{\rho}
\right] {\alpha}_{-s}^{\sigma} + \right. \]
\[ \left. {\alpha}_{m-n}^{\mu} {\alpha}_{s-m}^{\rho}
{\alpha}_n^{\nu} {\alpha}_{-s}^{\sigma} \right) \mid 0 \rangle =
\frac {1} {4} {\eta}_{\mu \nu} {\eta}_{\rho \sigma}
\sum\limits_{n=1}^{m-1} \sum\limits_{s=1}^{m-1} \]
\[ \left( \left[ {\alpha}_{m-n}^{\mu}, {\alpha}_{-s}^{\sigma} \right]
\left[ {\alpha}_n^{\nu}, {\alpha}_{s-m}^{\rho} \right] +
\left[ {\alpha}_{m-n}^{\nu}, {\alpha}_{s-m}^{\rho} \right]
\left[ {\alpha}_n^{\nu}, {\alpha}_{-s}^{\sigma} \right] \right) \]
\begin{equation}
A_m = \frac {D} {2} \sum\limits_{n=1}^{m-1} n(n-m) = \frac {D} {2}
\frac { m (m^2 -1)} {6}
\end{equation}

 Suppose now, for the sake of the argument, that we define the
vacuum to be annihilated by $ {\alpha}_{-m}^{\mu} $ with
$ m > 0 $
(Time inverted case) :

\begin{equation}
{\alpha}_{-m}^{\mu} \mid 0 \rangle =0 \;\;\;,\;\;\; L_{-m}
\mid 0 \rangle =0
\end{equation}

 The corresponding anomaly would be :

\[ A_m^{'} = \langle 0 \mid \left[ L_m, L_{-m} \right] \mid 0 \rangle =
- \langle 0 \mid L_{-m} L_m \mid 0 \rangle \]
\[ A_m^{'} =- \frac {1} {4} {\eta}_{\mu \nu} {\eta}_{\rho \sigma}
\sum\limits_{n=1}^{m-1} \sum\limits_{s=1}^{m-1} \left( \left[
{\alpha}_{s-m}^{\mu}, {\alpha}_n^{\sigma} \right] \right. \]
\[ \left. \left[ {\alpha}_{-s}^{\nu}, {\alpha}_{m-n}^{\rho} \right] +
\left[ {\alpha}_{s-m}^{\mu}, {\alpha}_{m-n}^{\rho} \right] \left[
{\alpha}_{-s}^{\nu}, {\alpha}_n^{\sigma} \right] \right) \]
\begin{equation}
A_m^{'} = - \frac {D} {2} \frac {m(m^2-1)} {6}
\end{equation}

 A mere change of sign with respect to (21).

 One can see the influence of the vacuum state (either (19) or (22))
on the value of the anomaly (resp.(21) or (23)). Thus,it is
understandable that the symmetrical identification expressed by eq.(14),
should lead to the disappearance of the anomaly.

 To check the last assertion we are going to consider again eq.(13) but
this time together with:

\begin{equation}
\langle 0 \mid {\alpha}_m^{\mu} {\alpha}_n^{\nu} \mid 0 \rangle =
m {\varepsilon}_m {\delta}_{m+n} {\eta}^{\mu \nu }
\end{equation}

 Eq.(24) covers all three possibilities, namely:

 I: The usual case,

\begin{eqnarray}
{\varepsilon}_m^I = & 1 \;\;\; if \;\;\; m>0  \nonumber \\
		    & 0 \;\;\; if \;\;\; m<0
\end{eqnarray}

 II: The inverted case,

\begin{eqnarray}
{\varepsilon}_m^{II} = & 0 \;\;\; if \;\;\; m>0 \nonumber \\
			 & 1 \;\;\; if \;\;\; m<0
\end{eqnarray}

 III: The symmetrical case,

\begin{equation}
{\varepsilon}_m^{III} = \frac {1} {2} \;\;\; (any \; m)
\end{equation}

 Let us now take a generic term in (13) (We will not write explicitly
the space-time indices):

\[ \langle 0 \mid {\alpha}_{m-n} {\alpha}_n {\alpha}_{-m-s}
{\alpha}_s \mid 0 \rangle = \left[ {\alpha}_n, {\alpha}_{-m-s}
\right] \cdot \]
\[ \cdot \langle 0 \mid {\alpha}_{m-n} {\alpha}_s \mid 0 \rangle +
\langle 0 \mid {\alpha}_{m-n} {\alpha}_{-m-s} {\alpha}_n
{\alpha}_s \mid 0 \rangle = \]
\[ n {\delta}_{n-m-s} {\varepsilon}_{m-n} (m-n) +
n {\delta}_{n+s} {\varepsilon}_{m-n} (m-n) + \]
\[ \langle 0 \mid {\alpha}_{m-n} {\alpha}_{-m-s} {\alpha}_s
{\alpha}_n \mid 0 \rangle = n(m-n) {\varepsilon}_{m-n}
{\delta}_{n-m-s} + \]
\[ n(m-n) {\varepsilon}_{m-n} {\delta}_{n+s} {\delta}_{-n-s} +
(m-n) s {\varepsilon}_s {\delta}_{n+s} +\]
\[ (m-n) (-m-s) {\varepsilon}_{-m-s} {\delta}_{n-m-s}
+ \langle 0 \mid {\alpha}_{-m-s} {\alpha}_s {\alpha}_{m-n}
{\alpha}_n \mid 0 \rangle \]

\begin{equation}
\langle 0 \mid \left[ {\alpha}_{m-n} {\alpha}_n , {\alpha}_{-m-s}
{\alpha}_s \right] \mid 0 \rangle = n(m-n) \left( {\varepsilon}_{m-n} -
{\varepsilon}_{-n} \right) \left( {\delta}_{n-m-s} + {\delta}_{n+s}
\right)
\end{equation}

 The last factor tells us that there is a possible contribution
only when s=-n or s=n-m.
 Let us examine the $ {\varepsilon} $ -factor. For case I (eq.(25)),
and $ m>0 $, there is no contribution for n outside the interval
$ 1 \leq n \leq m-1 $. With this information it is easy to
reproduce eq.(21).

 For case II (eq.(26)) and $ m>0 $,
the contributing interval is the same.
But in that interval we have $ {\varepsilon}_{m-n}^{II}=0 $, while in
case I we had $ {\varepsilon}_{-n}^I =0 $. This implies a change
of sign with respect to (21) as in eq.(23).

 Finally, for case III (eq.(27)), $ {\varepsilon}_m^{III} $ =
$ \frac {1} {2} $ , so that the $ \varepsilon $ -parenthesis is zero
and no anomaly is present.

 A similar calculation can be carried out for the "Lorentz anomaly"
\cite{tp14}. Its value depends on $ {\Delta}_m $ defined in ref.
\cite{tp14}, are all null for the symmetrical vacuum.

\section{ The ghosts }

 It is possible to follow a path-integral method for the treatment of
string motions. L.D.Faddeev and N.Popov procedure \cite{tp15}
\cite{tp16} to handle the gauge-fixing determinants leads to the use
of anti-commuting ghost fields, which should be considered together
with normal string states.

 The ghost action in the conformal gauge, can be written \cite{tp17}

\begin{equation}
S^c = \frac {1} {\pi} \int d^2 \sigma \left( c^{+} {\partial}_{-}
b_{+} + c^{-} {\partial}_{+} b_{-} \right)
\end{equation}

 where the ghost and anti-ghost fields obey the anticommutation relations

\begin{equation}
\lbrace b(\sigma, \tau), c({\sigma}^{'}, \tau) \rbrace = 2 \pi
\delta ( \sigma - {\sigma}^{'} )
\end{equation}

 We can now transform to normal mode coordinates.

\begin{equation}
c^{\pm}= \sum\limits_{- \infty}^{\infty} c_n e^{-in \left( \tau \pm
\sigma \right)}
\end{equation}
\begin{equation}
b_{\pm}= \sum\limits_{- \infty}^{\infty} b_n e^{-in \left( \tau \pm
\sigma \right)}
\end{equation}

 where $ c_n $ and $ b_n $ obey (from (30)):

\begin{equation}
\lbrace c_m, b_n \rbrace = {\delta}_{m+n}
\end{equation}

 The Fourier components of the world-sheet energy-momentum tensor
define the Virasoso generators \cite{tp17}:

\begin{equation}
L_m^c= \sum\limits_{- \infty}^{\infty} (m-n) b_{m+n} c_{-n}
\end{equation}

 They obey a Virasoro algebra with a possible central term:

\begin{equation}
\left[ L_m^c, L_n^c \right] = (m-n) L_{m+n}^c + A_n^c {\delta}_{m+n}
\end{equation}

 For the evaluation of $ A_m^c $ we will follow a method similar
to the one used in $ \S $ 4.

 We define three different vacuum states:

\begin{equation}
I \;\;\; b_n \mid 0 \rangle =0 \;\;,\;\; c_n \mid 0 \rangle =0
\;\;,\;\; n<0
\end{equation}
\begin{equation}
II \;\;\; b_{-n} \mid 0 \rangle =0 \;\;,\;\; c_{-n} \mid 0 \rangle =0
\;\;,\;\; n<0
\end{equation}
\begin{equation}
III \;\;\; \left( b_n c_{-n} - c_{-n} b_n \right) \mid 0 \rangle =0
\;\;,\;\;any\;n
\end{equation}

 The symmetrical ground state case III, has also been used to define
the vacuum corresponding to fermion fields with complex-mass parameter
\cite{tp19}.

 The contribution to the anomaly, produced by the ghosts, can now be
determined for the three cases.

 $ L_0^c $ is normal ordered so that:

\begin{equation}
A_m^c = \langle 0 \mid \left[ L_m^c, L_{-m}^c \right] \mid 0
\rangle
\end{equation}

 Also

\begin{equation}
\langle 0 \mid b_m c_n \mid 0 \rangle = {\varepsilon}_n
{\delta}_{n+m}
\end{equation}

 Where

\begin{eqnarray}
{\varepsilon}_n^I= & 1 \;\;,\;\; if \;\; n>0 \nonumber \\
		   & 0 \;\;,\;\; if \;\; n<0
\end{eqnarray}
\begin{eqnarray}
{\varepsilon}_n^{II}= & 0 \;\;,\;\; if \;\; n>0 \nonumber \\
			& 1 \;\;,\;\; if \;\; n<0
\end{eqnarray}
\begin{equation}
{\varepsilon}_n^{III} = \frac {1} {2} \;\;,\;\; any \; n
\end{equation}

 Now we take a generic term of the product $ L_m^c L_n^c $ :

\[ b_{m+r} c_{-r} b_{n+s} c_{-s} = b_{m+r} \lbrace c_{-r}, b_{n+s}
\rbrace c_{-s} - \]
\[ b_{m+r} b_{n+s} c_{-r} c_{-s} = {\delta}_{n+s-r} b_{m+r} c_{-s} - \]
\[ b_{n+s} \lbrace b_{m+r}, c_{-s} \rbrace c_{-r} + b_{n+s}
c_{-s} b_{m+r} c_{-r} \]

 Leading to:

\[ \langle 0 \mid \left[ b_{m+r} c_{-r}, b_{n+s} c_{-s} \right]
\mid 0 \rangle = {\delta}_{m+n} {\delta}_{m+r-s} \left(
{\varepsilon}_{-s} - {\varepsilon}_{-r} \right) \]

 i.e.:

\[ A_m^c = \langle 0 \mid \left[ L_m^c, L_{-m}^c \right] \mid 0
\rangle = \]
\begin{equation}
\sum\limits_s (m-s)(2m+s) \left( {\varepsilon}_{-m-s} - {\varepsilon}_{-s}
\right)
\end{equation}

 Again, we can see from (44) that the anomaly changes sign from I to
II, and disappears in case III.

\section {Discussion}

 In this paper we wanted to emphasize that the anomalies are
consequences of the asymmetry of the ground state with respect to
the lowering ( $ \alpha $ ) and raising ( $ {\alpha}^{+} $ )
operators. As a result, it follows that the choice of the vacuum
for the string determines the anomaly of the Virasoro algebra.
When the symmetry is restored, the anomaly disappears.

 As a matter of facts, we face three typical situations. First
(and most important), by succesive applications of $ \alpha $ we
arrive at a state $ \mid 0 \rangle $ such that $ \alpha \mid 0
\rangle $ $ =0 $ . Second, the inverted case where by succesive
applications of $ {\alpha}^{+} $ we arrive at a "ceiling state"
$ {\alpha}^{+} $ $ \mid 0 \rangle $ $ =0 $ . Third, no matter how
many times we multiply by $ {\alpha} $ or $ {\alpha}^{+} $, we
never annihilate a state. Those are the cases examined in $ \S $
3 and $ \S $ 4. They lead to different matrix representations
of $ {\alpha} $ and $ {\alpha}^{+} $ . Also they imply, of course,
different propagators for the fields that obey the oscillator
differential equation. The first case corresponds to the usual
Klein-Gordon equation and the associated Feynman propagator.
The second case takes place if we adopt the time inverted solution
and the complex-conjugate of the Feynman propagator. The third
alternative occurs when the field can not be observed asymptotically
(propagating freely). Such is the case for example, when the
Klein-Gordon equation has a complex-mass parameter and the free
solution blows-up asymptotically. The associated propagator is half
advanced and half retarded ( See ref. \cite{tp7} ).

 In the latter case, as the negative normed states can not be
excited, they can be considered to be auxiliary constructions
that may latter be discarded. Only the propagators are needed
for the subsequent developments of the theory.

 Sometimes doubts are raised about the costumary identification of
the vacuum ( See ref. \cite{tp11} ). Also, the asymptotic requirements
leading to the Feynman propagator do not seem to be compelling
for the finite string. For all those facts we think that a reasonable
alternative to the usual identification is provided by the
symmetrical vacuum state. Furthermore, in this way we obtain a
non-anomalous theory in any number of dimensions, in particular in
D=4. Consequently, we think that the possibilities opened up by
the adoption of the symmetrical vacuum are worth exploring.

\section*{Acknowledgments}

 We are grateful to O. Civitarese, A. L. De Paoli, J. D. Edelstein,
 H. Falomir, C. M. Naon, L. E. Oxman, M. Santangelo and
 H. Vucetich for helpful discussions.

\pagebreak

\end{document}